\documentclass[aps,prb,twocolumn,superscriptaddress,showpacs,floatfix]{revtex4}
\usepackage{graphicx}
\usepackage{amsmath}
\usepackage{amssymb}
\usepackage{url}

\newcommand{\D}{\ensuremath{{\mathrm{d}}}}

\graphicspath{{.}{../}{./eps/}}

\begin{document}
\title{Distribution of the local density of states as a criterion for
Anderson localization: Numerically exact results for various lattices in dimensions $D=2$
and $3$
}

\author{Gerald Schubert}
\affiliation{Regionales Rechenzentrum Erlangen,
Friedrich-Alexander-Universit\"at
  Erlangen-N\"urnberg, 91058 Erlangen, Germany}
\author{Jens Schleede}
\affiliation{Institut f\"ur Physik, Ernst-Moritz-Arndt Universit\"at
  Greifswald, 17487 Greifswald, Germany}
\author{Krzysztof Byczuk}
\affiliation{Institute of Theoretical Physics, University of Warsaw, 00-681
Warsaw, Poland}
\author{Holger Fehske}
\affiliation{Institut f\"ur Physik, Ernst-Moritz-Arndt Universit\"at
  Greifswald, 17487 Greifswald, Germany}
\author{Dieter Vollhardt}
\affiliation{Theoretical Physics III, Center for Electronic
  Correlations and Magnetism, Institute for Physics, University of
  Augsburg, 86135 Augsburg,
  Germany}

\date{\today}

\begin{abstract}
Numerical approaches to Anderson localization face the problem of having to treat
large localization lengths while being restricted
to finite system sizes.
We show that by finite-size scaling of the probability distribution of the local
density of states (LDOS) this long-standing problem can be overcome.
To this end we reexamine the approach, propose numerical refinements, and apply it
to study the dependence of the distribution of the LDOS on the dimensionality
and coordination number of the lattice. Particular attention is given to the graphene lattice.
We show that the system-size dependence of the LDOS distribution is indeed an
unambiguous sign of Anderson localization, irrespective of the dimension
and lattice structure.
The numerically exact LDOS data obtained by us agree with a
log-normal distribution over up to ten orders of magnitude
and thereby fulfill a nontrivial symmetry relation previously derived for 
the non-linear $\sigma$-model.

\end{abstract}



\maketitle

\section{Introduction}

The Anderson transition from extended to localized states in disordered systems
has been a subject of intense theoretical and numerical
investigations in the past~\cite{An58,LR85,VW92,KM93b,Evers08}, and is also of great current
interest~\cite{Lagendijk09,Aspect09,Science10,Lewenstein10}.
In particular, with the discovery of graphene~\cite{NGMJZDGF04} a truly
two-dimensional (2D)
system is now available in which localization effects can be
studied experimentally by direct observation of the local density of
states (LDOS) in scanning tunneling microscope
experiments~\cite{NKF09,NKMYF06,MKMW03}.
%


%
For infinite 2D systems (e.g.,  graphene) the one-parameter scaling theory
predicts electron localization to occur for arbitrary strengths of the Anderson disorder~\cite{AALR79}, with an exponentially large localization length for weak disorder.
In finite 2D systems, e.g., in currently available graphene flakes or potential
 graphene nano-devices
%
 extended states may thus exist if the sample is smaller than the localization length~\cite{SSF09}.

Early numerical work on localization concentrated on the calculation of the
averaged
localization length or the inverse participation
ratio of single-particle wave functions~\cite{LT74,We80}.
Quite generally investigations of disordered systems require the study of {\em distributions} of
local quantities. This makes the application of statistical methods necessary~\cite{An58,AAT73} to compute, for example,
%
 the distribution of squared wave function amplitudes~\cite{FE95},
participation ratios~\cite{Mi00}, or the LDOS~\cite{MF94b,Mi96,BAF04,AF05,AF08a}.
On the analytical side, great insight into disordered systems
was gained by the random matrix theory~\cite{Me91}.
Here the supersymmetric non-linear $\sigma$-model
formalism~\cite{Ef83} allows one to establish a link between
random matrices and Anderson-type disorder models.
Thereby the localization properties of a system are traced back to
global symmetry properties such as time reversal invariance.
Using the transfer-matrix technique, the $\sigma$-model may be solved
exactly for quasi-1D geometries~\cite{FM93}.
In higher dimensions, only approximate results were obtained so far, e.g., by employing
perturbative renormalization group~\cite{AKL91} or saddle
point-integration methods~\cite{MK95}.

In view of the predicated universality of the results of the random matrix theory
it is worthwhile to apply unbiased
numerical schemes to check the results and perhaps detect possible deviations.
To extract meaningful results from numerical studies
the data clearly need to be supplemented by a careful finite-size analysis.
In this way, the scaling properties of the distribution of the
squared wave function amplitude  were recently used to
determine the multifractal spectrum
at the Anderson transition~\cite{RVR09}.

While in principle all local quantities are equally suited to
describe localization properties in terms of their distribution,
for practical use the LDOS seems to be the most favorable.
This is due to its accessibility by means of highly efficient numerical
computation schemes, such as the Kernel Polynomial Method~(KPM)~\cite{WWAF06}.
Within the KPM the LDOS of any lattice site can be calculated
to, de facto, arbitrary precision without an explicit
diagonalization of the Hamiltonian.
A further advantage of the LDOS is the fact, that this single-particle quantity remains meaningful even in the presence of
interactions~\cite{Dobrosavljevic97,BF02,Dobrosavljevic03,Byczuk05+09,atkinson,Byczuk10},
 and because of its direct measurability by scanning tunneling microscopy~\cite{NKF09,NKMYF06,MKMW03,Science10}.

In this paper we revisit the local distribution approach for the LDOS,
propose further technical refinements of the KPM, and comment on potential
pitfalls concerning its application. In particular, we investigate the influence of the dimension and the coordination number of the underlying lattice
on the  LDOS distribution.
The results are contrasted with the analytically predicted
log-normal shape of the LDOS distribution~\cite{Mi00}.

\section{Model and method}

We consider the Anderson model of disorder,
\begin{equation}
  \label{model}
  {H} =
- t \sum\limits_{\langle ij \rangle}
  \bigl({c}_i^{\dag} {c}_j^{} + \text{H.c.}\bigr) + \sum\limits_{i} \epsilon_i {c}_i^{\dag} {c}_i^{} ,
\end{equation}
where the operators ${c}_i^{\dag}$ (${c}_i^{}$) create
(annihilate) an electron in a Wannier state centered at site $i$.
In~(\ref{model}), $t$ denotes the nearest-neighbor transfer integral.
Depending on the dimensionality ($D=2$ or $D=3$) and lattice structure (e.g., honeycomb,
hyper-cubic, triangular), the number of nearest neighbors varies from
three to six. The on-site potentials $\epsilon_i$ are drawn from
a uniform probability distribution (box distribution)
\begin{equation}\label{Kasten}
   p[\epsilon_i] = \frac{1}{\gamma}
   \;\theta\left(\frac{\gamma}{2}-|\epsilon_i|\right)\;.
\end{equation}
The parameter $\gamma$ measures the strength of the disorder.
Each different sample corresponds to one particular realization $\{\epsilon_i\}$.

\begin{figure}
     \centering\includegraphics[width=0.96\linewidth,clip]{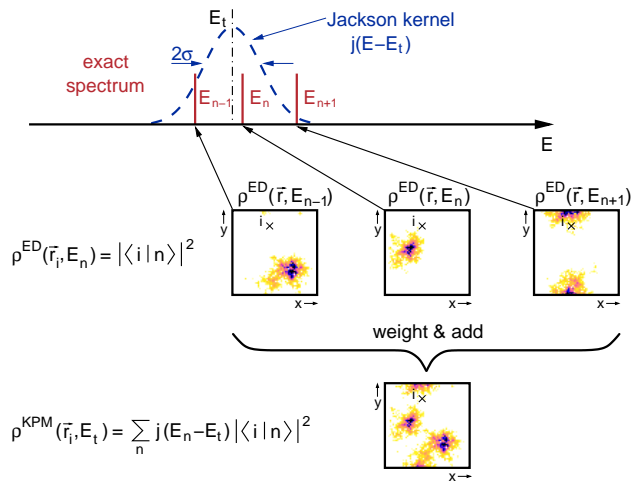}
     \caption{(Color online)
       Relation between the LDOS as calculated by ED and KPM.
       For a single eigenstate $|n\rangle$, the LDOS at site $\vec{r}_i$
       and energy $E_n$ is simply the squared amplitude of the wave function
       at this site, $\rho_i^{\text{ED}}(E_n) =
       \rho^{\text{ED}}(\vec{r}_i,E_n) = |\langle i|n\rangle|^2$.
       The squares schematically depict the
       LDOS values  on all lattice sites for three different eigenstates, using the same color map as in
       Fig.~\ref{fig:Snapshot_Graphite}. The cross indicates an arbitrary site
       $i$ of the sample.
       In contrast, the KPM assembles contributions from all eigenstates
in the
       energetic vicinity of the target energy $E_{\text{t}}$.
       Here the width $\sigma$ of the Jackson kernel $j(E-E_{\text{t}})$
       determines both the number of involved states and their weight in the
       superposition, $\rho_i^{\text{KPM}}(E_{\text{t}}) = \sum_n
       j(E_n-E_{\text{t}})|\langle i|n\rangle|^2$.
     }
     \label{fig:sketch_LDOS_KPM}
\end{figure}

Following Anderson~\cite{An58}, we consider the LDOS at lattice site $i$,
\begin{equation} \label{LDOS}
  \rho_i(E) = \sum\limits_{n=1}^{N}
  | \langle i | n \rangle |^2\, \delta(E-E_n)\;,
\end{equation}
where $|i\rangle = c^{\dag}_i |0\rangle$, and $|n\rangle$
is a single-electron eigenstate of $H$ with energy $E_n$,
 where $N$ denotes the number of sites and we assume periodic boundary
 conditions.
The LDOS can be determined very efficiently by the KPM.
The KPM is, in principle, an exact diagonalization~(ED) approach,
which is based on the expansion of the rescaled Hamiltonian
into a finite series of Chebyshev polynomials~\cite{WWAF06,WF08} .
Calculating the expansion coefficients  requires
only sparse matrix-vector multiplications with $H$.
In order to avoid Gibbs oscillations arising from the finite
expansion order, the exact spectrum is convoluted
with the strictly positive Jackson kernel.
In consequence, each $\delta$-peak is approximated by an almost
Gaussian of width $\sigma$, which depends on the order $M$ of the
polynomial approximation~\cite{WF08}.
By definition, the Gaussian is reduced to $e^{-1/2}\approx0.61$ of
the peak value at distance $\sigma$.
In the following, we will refer to $\sigma$ shortly as width of the
Jackson kernel.
Due to the peak broadening, the calculated LDOS at a
given energy, $E_{\text{t}}$, also contains contributions
from states with nearby energies (see Fig.~\ref{fig:sketch_LDOS_KPM}).
At first glance, this seems to be a drawback, preventing us from
resolving the localization properties of a single state.
However, the characteristics of one state for one realization of
disorder are irrelevant due to the statistical
nature of the Hamiltonian.
Instead, an ensemble of disorder realizations has to be
considered.
In this respect, the finite energy resolution is even helpful
since the exact eigenenergies vary from one random sample to the
next.
Hence the finite width of the Jackson kernel
ensures the inclusion of at least
some states with energies close to $E_{\text{t}}$.

Determining the localization properties of a system for given parameters
$(\gamma, E)$ by the local distribution approach is a two-step procedure.
In a first step, we accumulate the probability distribution of the LDOS at
energy $E$, $f[\rho_i(E)]$, within a histogram.
Therefore, we probe many different lattice sites $i$ and
realizations of disorder $\{\epsilon_i\}$.
By normalizing the LDOS to its mean value
$\rho_{\text{me}}(E) = \langle \rho_{i}(E)\rangle$, i.e.,
$\tilde\rho_{i} = \rho_i / \rho_{\text{me}}$,
we can pin down the characteristics of $f[\tilde\rho_i(E)]$ in terms of its shape. For this we note that the distributions $f[\tilde\rho_i(E)]$ match
with respect to norm and mean value for all $\gamma$ and $E$.
Since $\tilde\rho_i(E)$ varies over several orders of magnitude
for a disordered system, a logarithmic partitioning of the histogram
bins is advantageous. Equivalently, we can consider the
assembly of $\tilde f[\ln\tilde\rho_i(E)]$,
defined by the differentials
\begin{equation}
  f[\tilde\rho_i(E)] d[\tilde\rho_i(E)] =
  \tilde f[\ln\tilde\rho_i(E)]d[\ln\tilde\rho_i(E)]\,.
\end{equation}
\label{eq:ftilde}
In order to minimize the statistical fluctuations, the width of the
histogram bins $\xi$ has to be adapted to the calculated number of
LDOS values, $K$.
Following Scott~\cite{Sc79}, we use $\xi=3.49\sigma_0 K^{-1/3}$,
where $\sigma_{0}$ is the width of a Gaussian fit to
$\tilde f[\ln\tilde\rho_i(E)]$.
The logarithmic representation, $\tilde f[\ln\tilde\rho_i(E)]$,
enables a single comparison with the suggested
log-normal distribution~\cite{Mi00}:
\begin{equation}
  \phi_{0}(\tilde\rho_i) = \frac{1}{\sqrt{2\pi\sigma_0^2}}
  \frac{1}{\tilde\rho_i}
  \exp\left[-\frac{\left(\ln\tilde\rho_i-\mu\right)^2}
    {2\sigma_0^2}\right]\;.
  \label{log-normal}
\end{equation}
We note that in transforming $f\to\tilde{f}$ a Jacobian appears,
and the log-normal distribution on a linear scale acquires
an additional factor $1/\tilde\rho_i(E)$ as compared
to a Gaussian distribution on the logarithmic scale.
In consideration of the fact that the mean value and the norm of
$f[\tilde\rho_i(E)]$ are unity, the same should hold
for the distribution $\phi_{0}(\tilde\rho_i)$, \eqref{log-normal}.
This can be ensured by requiring $2\mu=-\sigma_0^2$ in Eq.~\eqref{log-normal},
thereby reducing the number of free fit-parameters to a single one.
Due to this additional requirement, $\phi_{0}(\tilde\rho_i)$ fulfills the
symmetry relation
\begin{equation}
  \tilde\rho_i^{3} \phi_{0}(\tilde\rho_i)=\phi_{0}(\tilde\rho_i^{-1}) \,,
  \label{EQ:R1}
\end{equation}
which was first derived for the non-linear $\sigma$-model~\cite{MF94,SSF05},
and which was recently shown to lead to the  prediction of a nontrivial symmetry
satisfied by multifractal exponents~\cite{MFME06}.
As any microscopic disordered system close to the transition point is equivalent to the
non-linear $\sigma$-model,
this relation should be valid also for the Anderson model~\eqref{model}.
A more detailed discussion of this issue is given in
Appendix A. 

The second, decisive, step of the local distribution approach is
the finite-size scaling of the obtained energy-dependent LDOS distributions,
which finally allows one to address the problem of Anderson localization.
Keeping the number of states within the Jackson kernel, $N_k$,
fixed, localized and extended states behave differently upon
increasing $N$:
For extended states, the normalization of $\tilde\rho_i$ to
$\rho_{\text{me}}$  ensures that $f[\tilde\rho_i(E)]$ is
independent of the system size.
In contrast, despite fixed norm and mean, localized states are
characterized by a shift of $f[\tilde\rho_i(E)]$ towards zero
as the system size increases.
In the theoretical limit of infinite system size, the
distribution will become singular at zero.
Additional, more technical details of the LDOS approach are given
in Appendix B.


\section{Results and Discussion}
\subsection{Three-dimensional lattices}
\subsubsection{Cubic lattice}

\begin{figure}
  \centering\includegraphics[width=0.96\linewidth,clip]{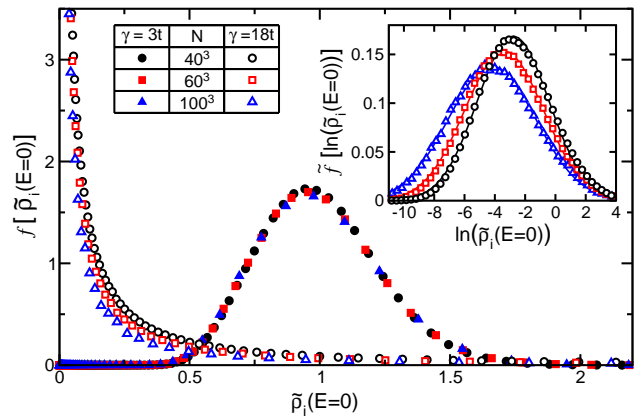}
  \caption{(Color online)
    Probability distribution of the normalized LDOS for
    states in the band center of a  3D cubic lattice.
     To ensure
    a proper statistics, the LDOS values of $K=2\times10^{6},
    4\times10^{5},8\times10^{4}$
    different sites and disorder realizations were calculated for
    $N=40^3,60^3,100^3$.
    The KPM resolution was adapted to contain $N_k=140$
    states within the kernel, irrespective of
    the system size. In the inset $\tilde f[\ln\tilde\rho_i]$ is given for the
    localized case ($\gamma=18t$) by symbols together with a
    log-normal fit to the data (solid lines).
        }
  \label{fig:3d_cubic}
\end{figure}

In order to illustrate the power and reliability of the local distribution
approach, let us first consider the 3D cubic lattice.
Here extended states exist for $\gamma<\gamma_c\approx16.5t$, whereas
for $\gamma >\gamma_c$ all states are localized~\cite{KM93b}.
In Fig.~\ref{fig:3d_cubic} the different scaling behaviors of
$f[\tilde\rho_i(E=0)]$ with increasing $N$ can be readily observed:
(i) For weak disorder ($\gamma=3t$) the distribution is completely
unaffected upon increasing the system size from $N=40^3$ to $N=100^3$.
The almost Gaussian shape of $f[\tilde\rho_i(E=0)]$ reflects the random
LDOS fluctuations from one site to another and the absence of any
superimposed global structure. Note that the $N$-independent
width of the distribution
(between 50\% and 150\% of $\rho_{\text{me}}$) is due
the fixed $N_k$ scaling and its
normalization to $\rho_{\text{me}}$. Of course, concerning
absolute values it narrows with increasing $N$ because of the normalization of
the individual eigenstates.
(ii) For strong disorder ($\gamma=18t$) the distribution
markedly depends on $N$.
Its shape distinctly differs from  a Gaussian distribution
 and closely resembles a
log-normal distribution (see inset of Fig.~\ref{fig:3d_cubic}).
The maximum shifts by an order of magnitude when going
from $N=40^3$ to $N=100^3$.
The LDOS data and their log-normal fit  are seen to match
perfectly over several orders of magnitude.
Here it should be emphasized that a distinction
between localized and extended
states solely based on the shape of the distribution at
fixed system size is not meaningful---only the behavior of
the distribution upon finite-size scaling is relevant~\cite{SWWF05}.
It is known that the distribution of the LDOS of extended states can be
described equally well by a log-normal distribution.
Both distributions agree for lattices without disorder, and
are hardly distinguishable as long as mean and most probable
value do not differ significantly.

\subsubsection{Honeycomb lattice}

\begin{figure}
     \centering\includegraphics[width=0.96\linewidth,clip]{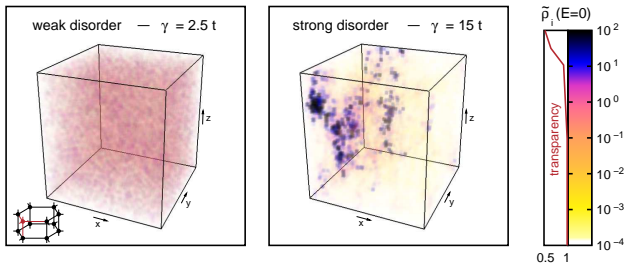}
     \caption{(Color online)
       Spatially resolved LDOS of the eigenstate
       with energy closest to $E=0$  for a particular disorder realization
       of the Anderson model.
       We contrast the results obtained for
       weak disorder ($\gamma=2.5t$), where the particle is spread over the
       whole sample (left-hand panel), with those for strong
       disorder ($\gamma=15t$), where the
       particle is confined to a finite region (right-hand panel).
       Data were obtained by ED for the 3D
       honeycomb lattice with $N=32^3$ sites.
       Each lattice point is represented by a small cube, with color
       corresponding to the squared amplitude of the wave function.
       In order to permit a view into the crystal, in addition we used
       the transparency gradient given on the left of the color bar.
     }
     \label{fig:Snapshot_Graphite}
\end{figure}

In addition  to the discussion of the LDOS distribution, we now
show how the spatial structure of the LDOS differs
for a typical, but particular exact eigenstate of a certain sample in the
extended and localized regime,
respectively~(see Fig.~\ref{fig:Snapshot_Graphite}).
To this end we switch from the 3D cubic lattice to a 3D honeycomb lattice
sketched in the left panel of Fig.~\ref{fig:Snapshot_Graphite},
whereby the number of nearest-neighbor sites is reduced from six to five.
The unperturbed bandwidth for the 3D honeycomb
lattice without Anderson disorder is $W_0=10t$
instead of $W_0=12t$ for the cubic one.
In order to figure out how the same disorder strength influences the
localization properties for different coordination numbers,
we have to compare results for the same ratio  of $\gamma/W_0$ rather than
those for absolute $\gamma/t$ values.
%
To make contact with results shown in Fig.~\ref{fig:3d_cubic}, we
therefore use $\gamma=2.5t=0.25W_0$ and $\gamma=15t=1.5W_0$ in order
to model the weak and strong disorder case, respectively.
Both for extended and localized states we recover the
above quoted characteristics. Namely, for weak disorder we find
that the LDOS spreads over the whole lattice,
showing moderate amplitude fluctuations
without a global structure
(left panel of Fig.~\ref{fig:Snapshot_Graphite}).
For strong disorder there occurs a concentration of the LDOS on
a spatially restricted, filamentary subset of all lattice sites
(right panel of Fig.~\ref{fig:Snapshot_Graphite}).
However, since Fig.~\ref{fig:Snapshot_Graphite} shows particular
eigenstates, the LDOS variations are more pronounced than for the
superposed KPM data (cf. Appendix B,
Fig.~\ref{fig:Illustrate_influence_kernel}).
We note that the finite-size scaling of $f[\tilde\rho_i(E=0)]$ for
the 3D honeycomb
lattice (not shown) qualitatively agrees with the one presented in
Fig.~\ref{fig:3d_cubic}.
Quantitatively, for the heavily disordered 3D honeycomb lattice,
$\tilde f[\ln\tilde\rho_i(E=0)]$ is concentrated at smaller values
for all $N$, indicating a stronger localization, i.e., a smaller
localization length.
Despite the similarity of this lattice to graphite, our
results may not be transferred directly to this system.
For graphite, the layer stacking differs (Bernal stacking), and
the interlayer coupling is much weaker than the intralayer one.

\subsection{Two-dimensional lattices}

\subsubsection{Square lattice}

\begin{figure}
  \centering\includegraphics[width=0.96\linewidth,clip]{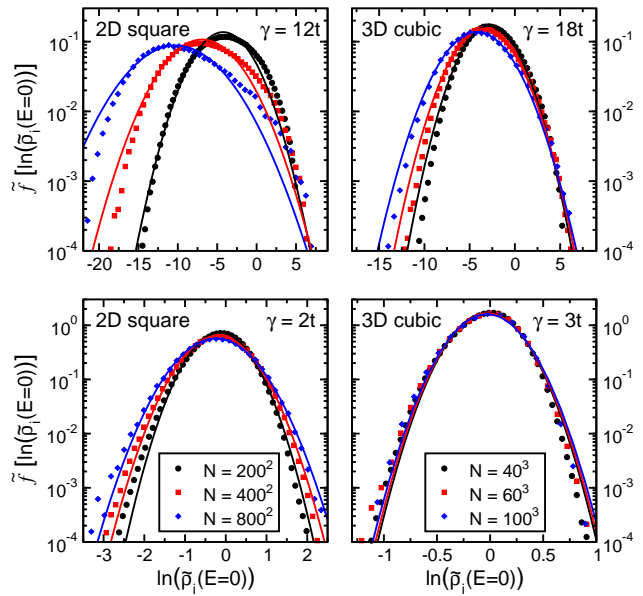}
  \caption{(Color online)
    Comparison of $\tilde f[\ln\left(\tilde\rho_i(E=0)\right)]$
    for  2D square and 3D cubic lattices. In both cases
    $\gamma= 0.25W_0$ and $\gamma=1.5W_0$. We used
    $N_k=140$ for
    all $N$ and assembled $K=2\times10^6,4\times10^{5},
    8\times10^{4}$ LDOS values
    for $N=40^3,60^3,100^3$ and $200^2,400^2,800^2$.
    The solid lines give fits to log-normal distributions.
  }
  \label{fig:fbar_2D3D}
\end{figure}

We now turn to the 2D case and compare the distribution of the LDOS
for the 3D cubic lattice with that for a 2D
square lattice (see Fig.~\ref{fig:fbar_2D3D}).
As predicted by the one-parameter scaling theory~\cite{AALR79},
arbitrarily weak disorder will cause localization in 2D.
However, the localization length can become so large that
the localized nature of the states can be easily missed
within a numerical approach.
Investigating a system with a localization length (much) larger than the
considered system size gives us the opportunity to demonstrate
the power of our method.
Again, we compare constant ratios $\gamma/W_0=0.25$ and
$\gamma/W_0=1.5$, respectively,
where the bandwidth for the 2D square lattice without disorder
is $W_0=8t$.
For strong disorder (upper panels in Fig.~\ref{fig:fbar_2D3D}),
the more pronounced shifting of
$\tilde f[\ln\left(\tilde\rho_i(E=0)\right)]$ indicates the even
stronger localization in 2D.
To speak of localization lengths is, however, dangerous due to
the ambiguous base for the finite-size scaling.
While the spectral properties are governed by the Hilbert
space dimension $N$, the localization length has to be compared
to the linear size of the system.
Quantitative statements about localization lengths are therefore limited
to systems of equal dimensionality.
The true challenge for any method is, of course, the weak disorder case (lower panels
in Fig.~\ref{fig:fbar_2D3D}), where an extended state (3D) has to be
distinguished from a localized one with large localization length (2D).
As already shown in Fig.~\ref{fig:3d_cubic} on a linear scale,
the 3D data show no dependence on $N$ at all on a double-logarithmic
scale, spanning several orders of magnitude.
By contrast, the $N$-dependence of the 2D data is obvious.
Although we do not see the pronounced shift of the maximum of the
distribution as in the strong disorder case, we observe a
systematic broadening
of the distribution.
Due to the constant norm and mean, this broadening can be
directly related to a shift of the maximum despite its faint
visibility.
By this, we may detect localized behavior even if the localization
length distinctly exceeds the system size.
A word of caution has to be added concerning the tails of the numerically
assembled distribution.
In an ensemble of $K$ LDOS values we cannot reliably expect to
find events outside the range $\xi \tilde f[\ln\tilde\rho_i] \gtrsim 1/K$
because of their low probability.
The prominent role of the distribution tails calls for maximizing
$K$ to extend their range of reliability.
Comparing the LDOS distributions to the corresponding log-normal fits,
the agreement is again remarkable.

\subsubsection{Triangular and honeycomb lattices}

\begin{figure}
  \centering\includegraphics[width=0.96\linewidth,clip]{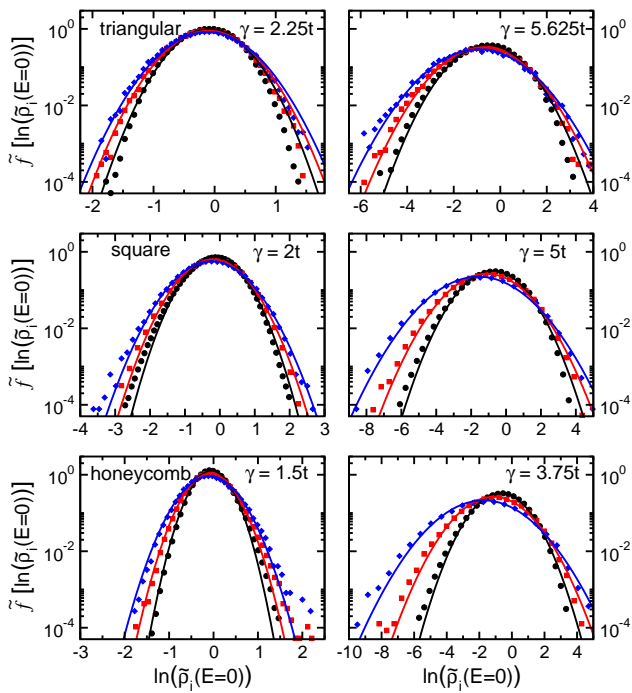}
  \caption{(Color online)
    Comparison of $\tilde f[\ln\left(\tilde\rho_i(E=0)\right)]$
    for different 2D lattices. We consider the cases of
   weak ($\gamma= 0.25W_0$) and
    moderate ($\gamma=0.625W_0$) disorder, with
    $N_k=140$.  For the triangular
    and square lattices the same symbols as in
    Fig.~\ref{fig:fbar_2D3D} for $N=200^2, 400^2, 800^2$ are used.
    For the honeycomb lattice we studied
    $N=192^2, 396^2, 792^2$ to match the correct boundary conditions and assembled $K=2\times10^{6},4\times10^{5},8\times10^{4}$ LDOS values (from small to
    large $N$).
  }
  \label{fig:2d_lattices}
\end{figure}

Finally we apply the local distribution approach to other 2D lattices and
address the influence of the coordination number on
the localization properties.
%
To this end we study the triangular and the
  honeycomb lattices.
In view of the current intense studies of graphene,
the latter is of great experimental and theoretical
relevance~\cite{NGMJZDGF04,CGPNG09}.
Calculating $\rho_i(E=0)$ for weakly disordered graphene requires
additional care due to the specifics of the electron dispersion
at and near the Dirac point.
For the ordered case we have $\rho_{\text{me}}\sim |E|$. At finite $N$,
with appropriate linear crystal size, four states
exist at $E=0$ surrounded by a finite-size gap and further stray
states.
Evaluating $\rho_{\text{me}}$ by means of the KPM, this value at $E=0$
will depend crucially on $\sigma$, i.e., on the broadening of the
vicinal peaks.
Furthermore, $\rho_{\text{me}}$ varies significantly on the
scale of the kernel width and can no longer be treated as constant
when calculating $N_k$.
As a consequence, the correct value of $N_k$ has to be
calculated self-consistently using \eqref{eq:N_k}
prior to the assembly of the LDOS histogram data.
According to the bipartite graphene lattice structure with three nearest
neighbors the bandwidth for the honeycomb lattice is $W_0=6t$.
Since the triangular lattice is non-bipartite, $\rho_{\text{me}}$ is
asymmetric with band edges in the ordered case at $E=-6t$
and $E=3t$.
Despite the six nearest neighbors of the triangular lattice, the
electronic bandwidth is $W_0=9t$ only, in contrast to the 3D cubic
lattice with $W_0=12t$.
In Fig.~\ref{fig:2d_lattices} we compare how weak
($\gamma=0.25W_0$) and moderate ($\gamma=0.625W_0$) disorder
influence electronic states on the above lattices.
In all six cases we observe clear indications for localization since
$\tilde f[\ln\left(\tilde\rho_i(E=0)\right)]$ either visibly shifts
on increasing $N$ (right column) or can be deduced from the systematic
tail broadening (left column).
The agreement with the corresponding log-normal fits also holds
equally well for all lattice types and both disorder strengths.
The strongest localization effects, i.e., the most pronounced shifting
of $\tilde f[\ln\left(\tilde\rho_i(E=0)\right)]$ can be observed for
the honeycomb lattice (see right column of Fig.~\ref{fig:2d_lattices}).
We attribute this to the smallest coordination number, a conclusion
that is supported by the less pronounced but clearly observable
localization for the triangular lattice.
%

\section{Conclusions}

By performing extensive, numerically exact calculations of the distribution of the
local density of states (LDOS) of disordered electrons on various two- and three-dimensional lattices we showed that the local distribution approach can reliably distinguish between localized and extended states.
A careful finite-size scaling analysis is crucial for the success of this method.
Localized states are identified by the broadening of the LDOS distribution
and its shift to zero for increasing system size.
Remarkably, a log-normal distribution perfectly fits the LDOS data
over up to ten orders of magnitude, depending on the dimensionality,
lattice structure and disorder strength, with only
minor systematic deviations for the outer tails of the distribution.
Close to the Anderson transition point the LDOS distribution 
corroborates a symmetry relation originally found for the
non-linear $\sigma$-model.
In contrast to localized states, the LDOS distribution for
extended states is system-size independent.

The stability of the Chebyshev expansion permits a
numerically exact calculation of the LDOS by means of 
the Kernel Polynomial Method,
the accuracy being limited only by machine precision.
Special care is required when calculating the LDOS at energies
where the mean DOS strongly varies.
Here one has to guarantee that the number of states within the
Jackson kernel is always kept constant in the finite-size scaling.
The symmetry relation of the LDOS can be used to improve 
the data analysis for Anderson disordered systems.

Our numerically exact results confirm the existence of a  localization
threshold in 3D as first predicted by Anderson~\cite{An58}, as well
as localization for any strength of disorder~\cite{AALR79} in 2D.
By increasing the coordination number of the lattice the influence of
disorder is reduced, such that comparable disorder strengths evoke
the shortest localization lengths for the graphene lattice.
Considering finite 2D, or quasi-1D graphene nanoribbons, the
localization length may nevertheless easily exceed the size of potential
nano-devices. In addition, boundary effects will become important.
As a result, charge transport will become possible despite the
existence of actually localized wave functions.
Concerning future investigations of the complex interplay between disorder and
electron-electron~\cite{LR85,BK94,KS04}
or electron-phonon interaction~\cite{BAF04} in finite, low-dimensional systems
the local density of states distribution approach shows great promise.

\section{Acknowledgments}

We thank A. Alvermann, F. X. Bronold, B. Shapiro, A. Wei{\ss}e,
G.~Wellein, and P. W\"olfle for stimulating discussions.
This work was funded by the Competence Network for Technical/Scientific
High-Performance Computing in Bavaria (KONWIHR II) and was supported
by TRR 80 and SPP 1459 of the Deutsche Forschungsgemeinschaft.
The numerical calculations were performed on the \mbox{HLRBII}, Leibniz
Supercomputing Center Munich, and on the tera-flop compute
cluster at the Institute of Physics, Greifs\-wald University and
Regional Computing Center Erlangen.

\appendix*

\section{A. Symmetry relation {\eqref{EQ:R1}} }

The log-normal distribution \eqref{log-normal} fulfills the
symmetry relation \eqref{EQ:R1} only if $2\mu=\sigma^2$.
Then, Eq.~\eqref{EQ:R1} can be proved
by simply completing the square in the exponent:
\begin{align}
  \tilde\rho_i^3 \phi_0(\tilde\rho_i) & = \frac{1}{\sqrt{2\pi\sigma_0^2}}
  \tilde\rho_i^2
  \exp\left[-\frac{(\ln\tilde\rho_i -\mu)^2}{2\sigma_0^2}\right] \nonumber\\
  & = \frac{1}{\sqrt{2\pi\sigma_0^2}}
  \tilde\rho_i
  \exp\left[-\frac{(\ln\tilde\rho_i -\mu)^2}{2\sigma_0^2} + \ln\tilde\rho_i
  \right]\nonumber\\
  & = \frac{1}{\sqrt{2\pi\sigma_0^2}}
  \tilde\rho_i
  \exp\left[ -\frac{(-\ln \tilde\rho_i -\mu)^2}{2\sigma_0^2}\right] \nonumber \\
  & = \phi_0(\tilde\rho_i^{-1})\,.
\end{align}
For any log-normal distribution that violates
$2\mu=\sigma^2$, the above procedure maps
$\phi_0(\tilde\rho_i)$ onto a different log-normal distribution,
which is shifted and rescaled.
A more intuitive understanding of this symmetry property can be
obtained by discussing it in terms of $x=\ln\tilde\rho_i$
instead of $\tilde\rho_i$.
On a logarithmic scale, Eq.~\eqref{EQ:R1} reads
\begin{equation}
  e^{x} \tilde\phi_0(x) = \tilde\phi_0(-x) \,,
  \label{EQ:R1_log}
\end{equation}
where $\tilde\phi_0(x)$ is a normal distribution.
Incorporating 
the $e^x$-factor into the exponential of  $\tilde\phi_0(x)$,
the position
of the maximum is shifted and the term that occurs by completing
the square gives an additional scaling factor.
For illustration consider the log-normal distributions
\begin{equation}
  \tilde{g}_{\pm}(x)  =
  \frac{1}{\sqrt{2\pi\sigma_0^2}}
  \frac{1}{1 + e^{\mp\alpha\sigma_0^2}}
  \exp\left[ -\frac{1}{2\sigma_0^2}
    \left(x  + \tfrac{\sigma^2_0}{2} \pm \alpha\sigma_0^2\right)^2\right],
\end{equation}
for which $ e^{x} \tilde{g}_{\pm}(x) = \tilde{g}_{\mp}(-x)$.
As a result, their sum
\begin{equation}
\tilde{\phi}_2(x)= \tilde{g}_{+} + \tilde{g}_{-}
\label{Sum}
\end{equation}
is invariant with respect to~\eqref{EQ:R1_log}.
The additional parameter $\alpha$ can be used in order to improve
the quality of the fit since the different prefactors of the
Gaussians in $\tilde{\phi}_2(x)$ also allow for an
adjustment to asymmetric data.
\begin{figure}
  \centering\includegraphics[width=0.96\linewidth,clip]{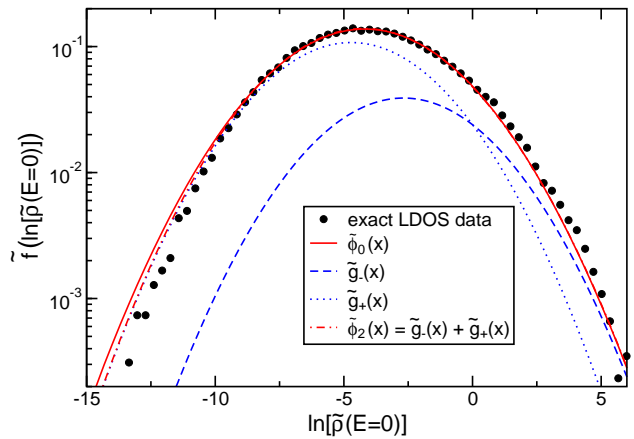}
  \caption{(Color online)
    Comparison of the quality of various fits to our LDOS data
     including the more
    complicated fit function, $\tilde{\phi}_2(x)$, \eqref{Sum}, comprising a
    sum of two log-normal distributions $g_{\pm}(x)$. Upon the
    symmetry operation~\eqref{EQ:R1_log} the function $g_{+}(x)$
    is mapped to $g_{-}(x)$ and vice versa. Note that
    $\tilde{\phi}_0(x)= \tilde{\phi}_2(\alpha=0,x)$.
    The LDOS data are taken from
    Fig.~\ref{fig:3d_cubic} and correspond to $\gamma=18t$, $N=100^3$.
  }
  \label{fig:lognormal_fit}
\end{figure}
Figure~\ref{fig:lognormal_fit} contrasts the best fits of
our LDOS data to $\tilde{\phi}_0(x)$  and $\tilde{\phi}_2(x)$.
Obviously, the improvement which can be achieved by the more
complicated $\tilde{\phi}_2(x)$ fit is not very significant.
Therefore, we restricted ourselves in the main part of the paper
to fits using $\tilde{\phi}_0(x)$.

\begin{figure}
  \centering\includegraphics[width=0.96\linewidth,clip]{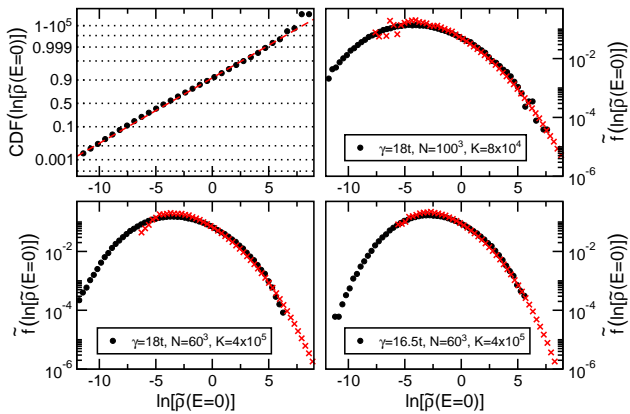}
  \caption{(Color online)
    Upper left panel: Log-normal probability plot for the LDOS distribution
    for $\gamma=18t$, $N=100^3$ taken from Fig.~\ref{fig:3d_cubic}.
     Only each second data point is depicted by a filled circle and compared
with the (red dashed) straight line.
      Other panels: Illustration to which extent the symmetry
      relation~\eqref{EQ:R1_log} holds for the exact LDOS data
      for different system sizes, number of realization,
      and disorder strengths. Calculated LDOS data are shown as
filled circles while crosses give the results
obtained by additionally exploiting the symmetry relation~\eqref{EQ:R1}.
The data in all panels were calculated using $N_k=140$. }
  \label{fig:lognormal_linear}
\end{figure}

The variation of the LDOS data over several orders of
magnitude makes it numerically challenging to single out
an optimal log-normal fit. In particular, it is not easy
to judge its quality by the eye.
Throughout the paper we employ the non-linear Levenberg-Marquardt
algorithm~\cite{PTVF92}. Thereby we used the pristine data
without any additional weighting (which might be introduced
to accentuate specific data regions, e.g., the distribution tails).
A well established graphical technique for testing the agreement of a
given data set with a (log-)normal distribution is the (log-)normal
probability plot~\cite{NIST03}.
Here the cumulative distribution function (CDF) is plotted and the ordinate
is scaled according to the inverse of the cumulative (log-)normal
distribution function.
Then, a (log-)normal distribution will show up as a straight line
(see upper left panel of Fig.~\ref{fig:lognormal_linear}).
This presentation underlines the almost perfect agreement of our
LDOS data with a (log-)normal distribution over the whole data range.
Deviations for large values of $\tilde\rho_i$
are due to the poor statistics of those rare events in our finite ensemble
of realizations.

We see that, by exploiting the symmetry~\eqref{EQ:R1} [or equivalently
Eq.~\eqref{EQ:R1_log}], any data analysis for disordered systems may 
profit from the statistically
more settled data near the maximum of the distribution in order
to enlarge the reliability of the data for very large values of $\rho_i$
(see other panels of Fig.~\ref{fig:lognormal_linear}).
Self-evidently the symmetry should be used in this direction only.
Mirroring the statistically poor data from the distribution tails
to regions near the distribution maximum is unprofitable.
Nevertheless, such a cross-check offers the
possibility to test the reliability of the numerical data,
where deviations may be attributed to bad statistics ($K$),
finite-size effects ($N$, $N_{\text{k}}$), or the fact
that we are too far away from the Anderson-transition point
($\gamma-\gamma_c$).
To illustrate the latter aspect, we determined the LDOS distribution
at $\gamma = 16.5t\simeq \gamma_c$, yielding---as compared to
$\gamma = 18t$ (for the same ($K$, $N$, $N_k$)---a better accordance with
the symmetry relation~\eqref{EQ:R1_log}.
Increasing $N$ clearly reduces the finite-size effects.
Therefore the data presented for $\gamma=18t$, $N=100^3$ are more reliable
than those for $N=60^3$ despite the fewer number of realizations $K$
in the latter case.

\section{B. Adaption of the KPM resolution}

The resolution of the original KPM is not uniform in the
whole energy interval.
For fixed expansion order the Jackson kernel becomes narrower
the closer it is located near the edges of the spectrum.
In order to allow for a clear distinction between resolution and
localization effects, a refinement of the KPM has been proposed in
Ref.~\onlinecite{SF09}, ensuring constant energy resolution over
the whole spectrum.
Still, this refinement does not account for one aspect that might become
important for systems with a strongly varying density of states (DOS),
like graphene:
If the DOS (and therefore the mean level spacing)
varies noticeably throughout the spectrum, an adjustment of
the kernel width $\sigma$ to $\rho_{\text{me}}$ is required.
Thus, by adapting $\sigma$, the same number of states within the Jackson kernel
\begin{equation}
  \label{eq:N_k}
  N_k(E_{\text{t}}) = N\int\limits_{E_{\text{t}}-\sigma(E_{\text{t}})}
  ^{E_{\text{t}}+\sigma(E_{\text{t}})}
  \rho_{\text{me}}(E)\D E
\end{equation}
is kept constant during the finite-size scaling for all energies $E_{\text{t}}$.
In many cases $\rho_{\text{me}}$ is approximately constant in the relevant
energy interval and $N_k\simeq 2\sigma N\rho_{\text{me}}$.
However, near the Dirac point in graphene the vanishing DOS requires
a self-consistent evaluation of \eqref{eq:N_k}.
Initial calculations of $\rho_{\text{me}}$ are performed
most efficiently by using delocalized random vectors in the KPM~\cite{WWAF06}.
This implicit averaging drastically improves the
statistical convergence of $\rho_{\text{me}}$, but is afterwards
unfeasible for the LDOS calculation itself.

\begin{figure}[b]
  \centering\includegraphics[width=0.96\linewidth,clip]{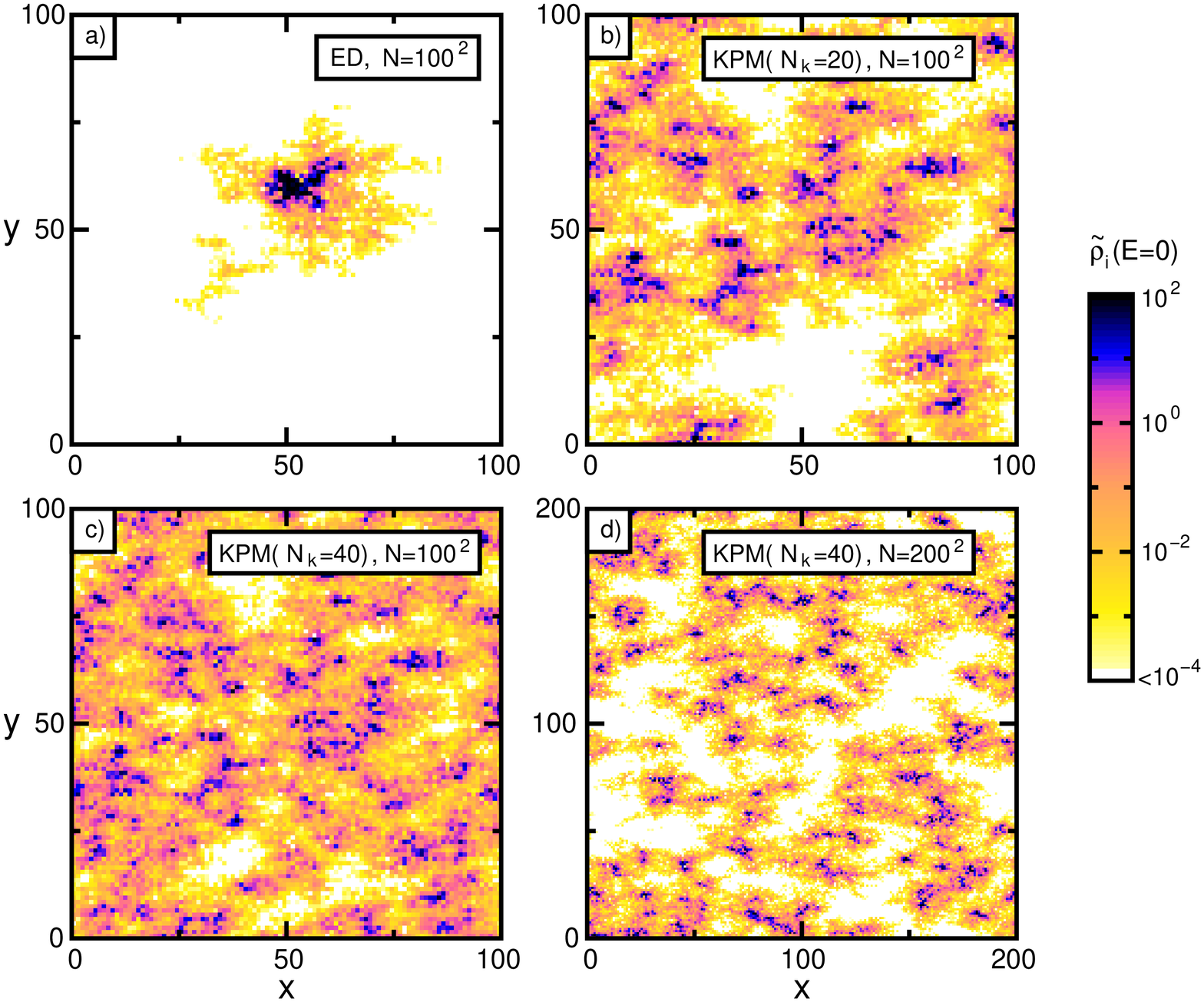}
  \centering\includegraphics[width=0.96\linewidth,clip]{fig8b.eps}
  \caption{(Color online) Influence of KPM resolution and
    system size on the LDOS and its distribution.
    Panel (a): LDOS of a single eigenstate in the band center
    for one realization of a strongly disordered ($\gamma=9t$)
    graphene lattice with $N=100^2$ sites. Panels (b)-(c):
    LDOS for the same realization
    calculated by KPM with different resolutions ($N_k=20,40$).
    Panel (d): LDOS for a larger system ($N=200^2$) at the same
    disorder strength and adapted KPM resolution such that
    $N_k=40$ is kept constant. Panel (e): Corresponding
    probability distributions of the (normalized) LDOS,
    $\tilde f[\ln(\tilde\rho_i(E=0))]$.
  }
  \label{fig:Illustrate_influence_kernel}
\end{figure}

A priori, there is no seeded value for $N_k$ to be
used in the finite-size scaling.
Only the limiting requirement of having some states within the
kernel for each realization of disorder has to be fulfilled.
In practice, one aims for a reasonable compromise between computational
costs (CPU time $\sim M \sim 1/\sigma\sim 1/N_k$) and energy resolution.
The concrete choice of $N_k$ has a significant impact on the LDOS
and its distribution (see Fig.~\ref{fig:Illustrate_influence_kernel}).
In the strongly localized regime individual eigenstates are
confined to a tiny fraction of all lattice sites
[Fig.~\ref{fig:Illustrate_influence_kernel}(a)].
Eigenstates with similar energies are localized at different
positions in space.
Due to the finite width of the Jackson kernel, the LDOS obtained
by KPM contains contributions of several of these states, depending
on $N_k$ [Fig.~\ref{fig:Illustrate_influence_kernel}(b) and
Fig.~\ref{fig:Illustrate_influence_kernel}(c)].
From Fig.~\ref{fig:Illustrate_influence_kernel}(c) it can be seen
that a too large value of $N_k$ masks the decay
of the individual eigenstates since
$\tilde\rho_i(E)$ becomes comparable
for most lattice sites.
Accessing the localization properties is, however, possible by
calculating the LDOS for a larger system using the same $N_k$
[Fig.~\ref{fig:Illustrate_influence_kernel}(d)] and comparing
both LDOS distributions [Fig.~\ref{fig:Illustrate_influence_kernel}(e)].
A shift of the LDOS distribution indicates Anderson localization.

Essentially, the properties of $\tilde f[\ln(\tilde\rho_i)]$ are
governed by the interplay of three parameters:
the localization length $\lambda$,
the system size $N$, and the number of states within the
Jackson kernel $N_k$.
Therefore, one might think of keeping $N$ fixed and investigate
the behavior of $\tilde f[\ln(\tilde\rho_i)]$ in dependence on
$N_k$ [curves corresponding to Fig.~\ref{fig:Illustrate_influence_kernel}(c)
and Fig.~\ref{fig:Illustrate_influence_kernel}(b)].
We use our initially proposed finite-size scaling
[Fig.~\ref{fig:Illustrate_influence_kernel}~(c) vs.
Fig.~\ref{fig:Illustrate_influence_kernel}~(d)]
since there the relevant ratio $\lambda/N$ is varied and
$N_k$ only enters as a parameter.
In comparison to the LDOS distribution from ED [curve corresponding to
Fig.~\ref{fig:Illustrate_influence_kernel}(a)], the LDOS distributions
obtained by KPM have the numerical advantage of a reduced width.
In consequence, limitations by machine precision will be less restrictive
for the KPM but might become an issue in ED for $\lambda\ll N^{1/D}$.

%


\begin{thebibliography}{32}
\expandafter\ifx\csname natexlab\endcsname\relax\def\natexlab#1{#1}\fi
\expandafter\ifx\csname bibnamefont\endcsname\relax
  \def\bibnamefont#1{#1}\fi
\expandafter\ifx\csname bibfnamefont\endcsname\relax
  \def\bibfnamefont#1{#1}\fi
\expandafter\ifx\csname citenamefont\endcsname\relax
  \def\citenamefont#1{#1}\fi
\expandafter\ifx\csname url\endcsname\relax
  \def\url#1{\texttt{#1}}\fi
\expandafter\ifx\csname urlprefix\endcsname\relax\def\urlprefix{URL }\fi
\providecommand{\bibinfo}[2]{#2}
\providecommand{\eprint}[2][]{\url{#2}}

\bibitem[{\citenamefont{Anderson}(1958)}]{An58}
\bibinfo{author}{\bibfnamefont{P.~W.} \bibnamefont{Anderson}},
  \bibinfo{journal}{Phys. Rev.} \textbf{\bibinfo{volume}{109}},
  \bibinfo{pages}{1492} (\bibinfo{year}{1958}).

\bibitem[{\citenamefont{Lee and Ramakrishnan}(1985)}]{LR85}
\bibinfo{author}{\bibfnamefont{P.~A.} \bibnamefont{Lee}} \bibnamefont{and}
  \bibinfo{author}{\bibfnamefont{T.~V.} \bibnamefont{Ramakrishnan}},
  \bibinfo{journal}{Rev. Mod. Phys.} \textbf{\bibinfo{volume}{57}},
  \bibinfo{pages}{287} (\bibinfo{year}{1985}).

\bibitem[{\citenamefont{Vollhardt and W\"olfle}(1992)}]{VW92}
\bibinfo{author}{\bibfnamefont{D.}~\bibnamefont{Vollhardt}} \bibnamefont{and}
  \bibinfo{author}{\bibfnamefont{P.}~\bibnamefont{W\"olfle}}, in
  \emph{\bibinfo{booktitle}{Electronic Phase Transitions}}, edited by
  \bibinfo{editor}{\bibfnamefont{W.}~\bibnamefont{Hanke}} \bibnamefont{and}
  \bibinfo{editor}{\bibfnamefont{Y.~V.} \bibnamefont{Kopaev}}
  (\bibinfo{publisher}{North Holland}, \bibinfo{address}{Amsterdam},
  \bibinfo{year}{1992}), p.~\bibinfo{pages}{1}.

\bibitem[{\citenamefont{Kramer and Mac~Kinnon}(1993)}]{KM93b}
\bibinfo{author}{\bibfnamefont{B.}~\bibnamefont{Kramer}} \bibnamefont{and}
  \bibinfo{author}{\bibfnamefont{A.}~\bibnamefont{Mac~Kinnon}},
  \bibinfo{journal}{Rep. Prog. Phys.} \textbf{\bibinfo{volume}{56}},
  \bibinfo{pages}{1469} (\bibinfo{year}{1993}).



\bibitem{Evers08} F. Evers, A.D. Mirlin, Rev. Mod. Phys. {\bf 80},
  1355 (2008).

\bibitem{Lagendijk09} A. Lagendijk, B. van Tiggelen, and
  D. S. Wiersma, Physics Today \textbf{62},
24  (2009).


\bibitem{Aspect09} A. Aspect and M. Inguscio, Physics Today \textbf{62},
30  (2009).


\bibitem{Science10} A. Richardella, P. Roushan, S. Mack, B. Zhou,
  D.A. Huse, D.D. Awschalom, and A. Yazdani, Science {\bf 327}, 665
  (2010).


\bibitem{Lewenstein10}  L. Sanchez-Palencia and M. Lewenstein, Nature
  Phys. \textbf{6}, 87 (2010).

\bibitem[{\citenamefont{Novoselov et~al.}(2004)\citenamefont{Novoselov, Geim,
  Morozov, Jiang, Zhang, Dubonos, Grigorieva, and Firsov}}]{NGMJZDGF04}
\bibinfo{author}{\bibfnamefont{K.~S.} \bibnamefont{Novoselov}},
  \bibinfo{author}{\bibfnamefont{A.~K.} \bibnamefont{Geim}},
  \bibinfo{author}{\bibfnamefont{S.~V.} \bibnamefont{Morozov}},
  \bibinfo{author}{\bibfnamefont{D.}~\bibnamefont{Jiang}},
  \bibinfo{author}{\bibfnamefont{Y.}~\bibnamefont{Zhang}},
  \bibinfo{author}{\bibfnamefont{S.~V.} \bibnamefont{Dubonos}},
  \bibinfo{author}{\bibfnamefont{I.~V.} \bibnamefont{Grigorieva}},
  \bibnamefont{and} \bibinfo{author}{\bibfnamefont{A.~A.}
  \bibnamefont{Firsov}}, \bibinfo{journal}{Science}
  \textbf{\bibinfo{volume}{306}}, \bibinfo{pages}{666} (\bibinfo{year}{2004}).

\bibitem[{\citenamefont{Niimi et~al.}(2009)\citenamefont{Niimi, Kambara, and
  Fukuyama}}]{NKF09}
\bibinfo{author}{\bibfnamefont{Y.}~\bibnamefont{Niimi}},
  \bibinfo{author}{\bibfnamefont{H.}~\bibnamefont{Kambara}}, \bibnamefont{and}
  \bibinfo{author}{\bibfnamefont{H.}~\bibnamefont{Fukuyama}},
  \bibinfo{journal}{Phys. Rev. Lett.} \textbf{\bibinfo{volume}{102}},
  \bibinfo{pages}{026803} (\bibinfo{year}{2009}).

\bibitem[{\citenamefont{Niimi et~al.}(2006)\citenamefont{Niimi, Kambara,
  Matsui, Yoshioka, and Fukuyama}}]{NKMYF06}
\bibinfo{author}{\bibfnamefont{Y.}~\bibnamefont{Niimi}},
  \bibinfo{author}{\bibfnamefont{H.}~\bibnamefont{Kambara}},
  \bibinfo{author}{\bibfnamefont{T.}~\bibnamefont{Matsui}},
  \bibinfo{author}{\bibfnamefont{D.}~\bibnamefont{Yoshioka}}, \bibnamefont{and}
  \bibinfo{author}{\bibfnamefont{H.}~\bibnamefont{Fukuyama}},
  \bibinfo{journal}{Phys. Rev. Lett.} \textbf{\bibinfo{volume}{97}},
  \bibinfo{pages}{236804} (\bibinfo{year}{2006}).

\bibitem[{\citenamefont{Morgenstern et~al.}(2003)\citenamefont{Morgenstern,
  Klijn, Meyer, and Wiesendanger}}]{MKMW03}
\bibinfo{author}{\bibfnamefont{M.}~\bibnamefont{Morgenstern}},
  \bibinfo{author}{\bibfnamefont{J.}~\bibnamefont{Klijn}},
  \bibinfo{author}{\bibfnamefont{C.}~\bibnamefont{Meyer}}, \bibnamefont{and}
  \bibinfo{author}{\bibfnamefont{R.}~\bibnamefont{Wiesendanger}},
  \bibinfo{journal}{Phys. Rev. Lett.} \textbf{\bibinfo{volume}{90}},
  \bibinfo{pages}{056804} (\bibinfo{year}{2003}).

\bibitem{SSF09}  G.~Schubert, J.~Schleede and H.~Fehske,
  Phys. Rev. B \textbf{79}, 235116 (2009).


\bibitem[{\citenamefont{Abrahams et~al.}(1979)\citenamefont{Abrahams, Anderson,
  Licciardello, and Ramakrishnan}}]{AALR79}
\bibinfo{author}{\bibfnamefont{E.}~\bibnamefont{Abrahams}},
  \bibinfo{author}{\bibfnamefont{P.~W.} \bibnamefont{Anderson}},
  \bibinfo{author}{\bibfnamefont{D.~C.} \bibnamefont{Licciardello}},
  \bibnamefont{and} \bibinfo{author}{\bibfnamefont{T.~V.}
  \bibnamefont{Ramakrishnan}}, \bibinfo{journal}{Phys. Rev. Lett.}
  \textbf{\bibinfo{volume}{42}}, \bibinfo{pages}{673} (\bibinfo{year}{1979}).

\bibitem[{\citenamefont{Last and Thouless}(1974)}]{LT74}
\bibinfo{author}{\bibfnamefont{B.~J.} \bibnamefont{Last}} \bibnamefont{and}
  \bibinfo{author}{\bibfnamefont{D.~J.} \bibnamefont{Thouless}},
  \bibinfo{journal}{J. Phys. C} \textbf{\bibinfo{volume}{7}},
  \bibinfo{pages}{699} (\bibinfo{year}{1974}).

\bibitem[{\citenamefont{Wegner}(1980)}]{We80}
\bibinfo{author}{\bibfnamefont{F.}~\bibnamefont{Wegner}}, \bibinfo{journal}{Z.
  Phys. B} \textbf{\bibinfo{volume}{36}}, \bibinfo{pages}{209}
  (\bibinfo{year}{1980}).

\bibitem[{\citenamefont{Abou-Chacra et~al.}(1973)\citenamefont{Abou-Chacra,
  Thouless, and Anderson}}]{AAT73}
\bibinfo{author}{\bibfnamefont{R.}~\bibnamefont{Abou-Chacra}},
  \bibinfo{author}{\bibfnamefont{D.~J.} \bibnamefont{Thouless}},
  \bibnamefont{and} \bibinfo{author}{\bibfnamefont{P.~W.}
  \bibnamefont{Anderson}}, \bibinfo{journal}{J. Phys. C}
  \textbf{\bibinfo{volume}{6}}, \bibinfo{pages}{1734} (\bibinfo{year}{1973}).

\bibitem[{\citenamefont{Fal\'{}ko and Efetov}(1995)}]{FE95}
\bibinfo{author}{\bibfnamefont{V.~I.} \bibnamefont{Fal\'{}ko}}
  \bibnamefont{and} \bibinfo{author}{\bibfnamefont{K.~B.}
  \bibnamefont{Efetov}}, \bibinfo{journal}{Phys. Rev. B}
  \textbf{\bibinfo{volume}{52}}, \bibinfo{pages}{17413} (\bibinfo{year}{1995}).

\bibitem[{\citenamefont{Mirlin}(2000)}]{Mi00}
\bibinfo{author}{\bibfnamefont{A.~D.} \bibnamefont{Mirlin}},
  \bibinfo{journal}{Physics Reports} \textbf{\bibinfo{volume}{326}},
  \bibinfo{pages}{259} (\bibinfo{year}{2000}).

\bibitem[{\citenamefont{Mirlin and Fyodorov}(1994)}]{MF94b}
\bibinfo{author}{\bibfnamefont{A.~D.} \bibnamefont{Mirlin}} \bibnamefont{and}
  \bibinfo{author}{\bibfnamefont{Y.~V.} \bibnamefont{Fyodorov}},
  \bibinfo{journal}{Phys. Rev. Lett.} \textbf{\bibinfo{volume}{72}},
  \bibinfo{pages}{526} (\bibinfo{year}{1994}).

\bibitem[{\citenamefont{Mirlin}(1996)}]{Mi96}
\bibinfo{author}{\bibfnamefont{A.~D.} \bibnamefont{Mirlin}},
  \bibinfo{journal}{Phys. Rev. B} \textbf{\bibinfo{volume}{53}},
  \bibinfo{pages}{1186} (\bibinfo{year}{1996}).

\bibitem[{\citenamefont{Bronold et~al.}(2004)\citenamefont{Bronold, Alvermann,
  and Fehske}}]{BAF04}
\bibinfo{author}{\bibfnamefont{F.~X.} \bibnamefont{Bronold}},
  \bibinfo{author}{\bibfnamefont{A.}~\bibnamefont{Alvermann}},
  \bibnamefont{and} \bibinfo{author}{\bibfnamefont{H.}~\bibnamefont{Fehske}},
  \bibinfo{journal}{Philos. Mag.} \textbf{\bibinfo{volume}{84}},
  \bibinfo{pages}{673} (\bibinfo{year}{2004}).

\bibitem[{\citenamefont{Alvermann and Fehske}(2005)}]{AF05}
\bibinfo{author}{\bibfnamefont{A.}~\bibnamefont{Alvermann}} \bibnamefont{and}
  \bibinfo{author}{\bibfnamefont{H.}~\bibnamefont{Fehske}},
  \bibinfo{journal}{Eur. Phys. J. B} \textbf{\bibinfo{volume}{48}},
  \bibinfo{pages}{295} (\bibinfo{year}{2005}).

\bibitem[{\citenamefont{Alvermann and Fehske}(2008)}]{AF08a}
\bibinfo{author}{\bibfnamefont{A.}~\bibnamefont{Alvermann}} \bibnamefont{and}
  \bibinfo{author}{\bibfnamefont{H.}~\bibnamefont{Fehske}},
  \bibinfo{journal}{Lecture Notes in Physics} \textbf{\bibinfo{volume}{739}},
  \bibinfo{pages}{505} (\bibinfo{year}{2008}).

\bibitem[{\citenamefont{Mehta}(1991)}]{Me91}
\bibinfo{author}{\bibfnamefont{M.~L.} \bibnamefont{Mehta}},
  \emph{\bibinfo{title}{Random Matrices}} (\bibinfo{publisher}{Academic Press},
  \bibinfo{address}{Boston}, \bibinfo{year}{1991}).

\bibitem[{\citenamefont{Efetov}(1983)}]{Ef83}
\bibinfo{author}{\bibfnamefont{K.~B.} \bibnamefont{Efetov}},
  \bibinfo{journal}{Adv. Phys.} \textbf{\bibinfo{volume}{32}},
  \bibinfo{pages}{53} (\bibinfo{year}{1983}).

\bibitem[{\citenamefont{Fyodorov and Mirlin}(1993)}]{FM93}
\bibinfo{author}{\bibfnamefont{Y.~V.} \bibnamefont{Fyodorov}} \bibnamefont{and}
  \bibinfo{author}{\bibfnamefont{A.~D.} \bibnamefont{Mirlin}},
  \bibinfo{journal}{Phys. Rev. Lett.} \textbf{\bibinfo{volume}{71}},
  \bibinfo{pages}{412} (\bibinfo{year}{1993}).

\bibitem[{\citenamefont{Altshuler et~al.}(1991)\citenamefont{Altshuler,
  Kravtsov, and Lerner}}]{AKL91}
\bibinfo{author}{\bibfnamefont{B.~L.} \bibnamefont{Altshuler}},
  \bibinfo{author}{\bibfnamefont{V.~E.} \bibnamefont{Kravtsov}},
  \bibnamefont{and} \bibinfo{author}{\bibfnamefont{I.~V.}
  \bibnamefont{Lerner}}, in \emph{\bibinfo{booktitle}{Mesoscopic Phenomena in
  Solids}}, edited by \bibinfo{editor}{\bibfnamefont{B.~L.}
  \bibnamefont{Altshuler}}, \bibinfo{editor}{\bibfnamefont{P.~A.}
  \bibnamefont{Lee}}, \bibnamefont{and} \bibinfo{editor}{\bibfnamefont{R.~A.}
  \bibnamefont{Webb}} (\bibinfo{publisher}{North-Holland},
  \bibinfo{address}{Amsterdam}, \bibinfo{year}{1991}), p. \bibinfo{pages}{449}.

\bibitem[{\citenamefont{Muzykantskii and Khmelnitskii}(1995)}]{MK95}
\bibinfo{author}{\bibfnamefont{B.~A.} \bibnamefont{Muzykantskii}}
  \bibnamefont{and} \bibinfo{author}{\bibfnamefont{D.~E.}
  \bibnamefont{Khmelnitskii}}, \bibinfo{journal}{Phys. Rev. B}
  \textbf{\bibinfo{volume}{51}}, \bibinfo{pages}{5480} (\bibinfo{year}{1995}).

\bibitem[{\citenamefont{Rodriguez et~al.}(2009)\citenamefont{Rodriguez,
  Vasquez, and R\"{o}mer}}]{RVR09}
\bibinfo{author}{\bibfnamefont{A.}~\bibnamefont{Rodriguez}},
  \bibinfo{author}{\bibfnamefont{L.~J.} \bibnamefont{Vasquez}},
  \bibnamefont{and} \bibinfo{author}{\bibfnamefont{R.~A.}
  \bibnamefont{R\"{o}mer}}, \bibinfo{journal}{Phys. Rev. Lett.}
  \textbf{\bibinfo{volume}{102}}, \bibinfo{pages}{106406}
  (\bibinfo{year}{2009}).

\bibitem[{\citenamefont{Wei{\ss}e et~al.}(2006)\citenamefont{Wei{\ss}e,
  Wellein, Alvermann, and Fehske}}]{WWAF06}
\bibinfo{author}{\bibfnamefont{A.}~\bibnamefont{Wei{\ss}e}},
  \bibinfo{author}{\bibfnamefont{G.}~\bibnamefont{Wellein}},
  \bibinfo{author}{\bibfnamefont{A.}~\bibnamefont{Alvermann}},
  \bibnamefont{and} \bibinfo{author}{\bibfnamefont{H.}~\bibnamefont{Fehske}},
  \bibinfo{journal}{Rev. Mod. Phys.} \textbf{\bibinfo{volume}{78}},
  \bibinfo{pages}{275} (\bibinfo{year}{2006}).

\bibitem{Dobrosavljevic97} V.~Dobrosavljevi\'c and G.~Kotliar,
  Phys. Rev. Lett. {\bf 78}, 3943 (1997); Phil. Trans. R. Soc. Lond. A \textbf{356}, 57 (1998).

\bibitem{BF02}  F.~X.~Bronold and H.~Fehske, Phys. Rev. B {\bf 66}, 073102 (2002).

\bibitem{Dobrosavljevic03} V.~Dobrosavljevi\'c, A.~A.~Pastor, and
  B.~K.~Nikoli\'c, Europhys. Lett. {\bf 62}, 76 (2003).

\bibitem{Byczuk05+09} K. Byczuk, W. Hofstetter, and D. Vollhardt,
  Phys. Rev. Lett. {\bf 94}, 056404 (2005); Phys. Rev. Lett. \textbf{102}, 146403 (2009).

\bibitem{atkinson} Yun Song, R. Wortis, and W. A. Atkinson,
  Phys. Rev. B {\bf 77}, 054202 (2008).

\bibitem{Byczuk10} K. Byczuk, W. Hofstetter, and D. Vollhardt,
to be published in \emph{Fifty Years of Anderson Localization}, edited by E. Abrahams (World Scientific, Singapore, 2010).


\bibitem[{\citenamefont{Wei{\ss}e and Fehske}(2008)}]{WF08}
\bibinfo{author}{\bibfnamefont{A.}~\bibnamefont{Wei{\ss}e}} \bibnamefont{and}
  \bibinfo{author}{\bibfnamefont{H.}~\bibnamefont{Fehske}},
  \bibinfo{journal}{Lecture Notes in Physics} \textbf{\bibinfo{volume}{739}},
  \bibinfo{pages}{545} (\bibinfo{year}{2008}).

\bibitem[{\citenamefont{Scott}(1979)}]{Sc79}
\bibinfo{author}{\bibfnamefont{D.~W.} \bibnamefont{Scott}},
  \bibinfo{journal}{Biometrika} \textbf{\bibinfo{volume}{66}},
  \bibinfo{pages}{605} (\bibinfo{year}{1979}).

\bibitem{MF94} A.~D.~Mirlin and Y.~V.~Fyodorov,
  J. Phys. I France. {\bf 4}, 655 (1994).

\bibitem{SSF05}
D. V. Savin, H.-J. Sommers, and Y. V. Fyodorov, Sov. Phys.
JETP Letters {\bf 82}, 544 (2005).

\bibitem{MFME06} A.~D.~Mirlin, Y.~V.~Fyodorov, A.~Mildenberger, and F. Evers
  Phys. Rev. Lett. {\bf 97}, 046803 (2006).


\bibitem[{\citenamefont{Schubert et~al.}(2005)\citenamefont{Schubert,
  Wei{\ss}e, Wellein, and Fehske}}]{SWWF05}
\bibinfo{author}{\bibfnamefont{G.}~\bibnamefont{Schubert}},
  \bibinfo{author}{\bibfnamefont{A.}~\bibnamefont{Wei{\ss}e}},
  \bibinfo{author}{\bibfnamefont{G.}~\bibnamefont{Wellein}}, \bibnamefont{and}
  \bibinfo{author}{\bibfnamefont{H.}~\bibnamefont{Fehske}}, in
  \emph{\bibinfo{booktitle}{High performance computing in science and
  engineering, Garching 2004}}, edited by
  \bibinfo{editor}{\bibfnamefont{A.}~\bibnamefont{Bode}} \bibnamefont{and}
  \bibinfo{editor}{\bibfnamefont{F.}~\bibnamefont{Durst}}
  (\bibinfo{publisher}{Springer-Verlag}, \bibinfo{address}{Heidelberg},
  \bibinfo{year}{2005}), pp. \bibinfo{pages}{237--249}.



\bibitem[{\citenamefont{{Castro Neto} et~al.}(2009)\citenamefont{{Castro Neto},
  Guinea, Peres, Novoselov, and Geim}}]{CGPNG09}
\bibinfo{author}{\bibfnamefont{A.~H.} \bibnamefont{{Castro Neto}}},
  \bibinfo{author}{\bibfnamefont{F.}~\bibnamefont{Guinea}},
  \bibinfo{author}{\bibfnamefont{N.~M.~R.} \bibnamefont{Peres}},
  \bibinfo{author}{\bibfnamefont{K.~S.} \bibnamefont{Novoselov}},
  \bibnamefont{and} \bibinfo{author}{\bibfnamefont{A.~K.} \bibnamefont{Geim}},
  \bibinfo{journal}{Rev. Mod. Phys.} \textbf{\bibinfo{volume}{81}}, \bibinfo{pages}{109} (\bibinfo{year}{2009}).



\bibitem{BK94}  D.~Belitz and T.~R.~Kirkpatrick,
  Rev. Mod. Phys. {\bf 66}, 261 (1994).

\bibitem{KS04} S.~V.~Kravchenko and M.~P.~Sarachik,
  Rep. Prog. Phys. {\bf 67}, 1 (2004).


%




\bibitem{PTVF92} W. H. Press and S. A. Teukolsky and W. T. Vetterling
                  and B. P. Flannery
  {\emph Numerical Recipes in C: The Art of Scientific
                  Computing},
  (1992, Cambridge University Press).


\bibitem{NIST03}NIST/SEMATECH e-Handbook of Statistical Methods, http://www.itl.nist.gov/div898/handbook/, 6/01/2003.

\bibitem[{\citenamefont{Schubert and Fehske}(2009)}]{SF09}
\bibinfo{author}{\bibfnamefont{G.}~\bibnamefont{Schubert}} \bibnamefont{and}
  \bibinfo{author}{\bibfnamefont{H.}~\bibnamefont{Fehske}},
  \bibinfo{journal}{Lecture Notes in Physics} \textbf{\bibinfo{volume}{762}},
  \bibinfo{pages}{135} (\bibinfo{year}{2009}).



\end{thebibliography}

\end{document}